\documentclass{aa}

\usepackage{graphicx}
\usepackage{txfonts}
\usepackage{natbib}
\bibpunct{(}{)}{;}{a}{}{,}
\usepackage[colorlinks=true, linkcolor=blue, citecolor=blue, urlcolor=blue]{hyperref}
\usepackage{array}
\usepackage{booktabs}
\usepackage{multirow}

\begin{document}

\titlerunning{CDDR test with clusters, Pantheon+, and DESI BAO}
\authorrunning{Hu et al.}

\title{A model-independent test of the cosmic distance-duality relation using galaxy clusters and Type Ia supernovae matched pairs}

\author{Jian Hu\inst{1}
\and
Yi Liu\inst{2}
\and
Jian-Ping Hu\inst{3}
\and
Zhongmu Li\inst{1}}

\institute{Institute of Astronomy and Information, Dali University, Dali 671003, People's Republic of China\\
\email{dg1626002@smail.nju.edu.cn}
\and
Department of Engineering, Dali University, Dali 671003, People's Republic of China\\
\email{lyly1974@hotmail.com}
\and
School of Astronomy and Space Science, Nanjing University, Nanjing 210093, People's Republic of China}

\date{Received ...; accepted ...}

\abstract
{The cosmic distance-duality relation (CDDR), $D_{L}/D_{A}(1+z)^{-2}=1$, is a fundamental premise in observational cosmology, linking the luminosity distance ($D_L$) and angular diameter distance ($D_A$). Any departure from this relation would point to new physics such as photon non-conservation, opacity, or non-metric gravity.}
{We perform a stringent, model-independent test of the CDDR using the latest observational data. We additionally assess the robustness of the matched-pair analysis against the choice of matching tolerance and supplement the baseline analysis with DESI BAO measurements. We aim to address the degeneracy between cosmic opacity and the source evolution of Type Ia supernovae (SNe~Ia) to provide robust constraints on the Etherington reciprocity theorem.}
{We utilize a matched sample of 38 galaxy clusters from the Bonamente et al. compilation and SNe~Ia from the Pantheon+ compilation. To avoid cosmological model dependence, we employ the matched pair technique, pairing $D_A$ and $D_L$ at identical redshifts. We adopt a joint likelihood analysis that simultaneously constrains the CDDR-violation parameter $\eta$ and a possible redshift evolution of the SNe~Ia absolute magnitude, parameterized as $M_{B}(z)=M_{0}+\varepsilon z$. We propagate the sub-covariance matrix extracted from the Pantheon+ compilation and implement a split-normal likelihood to preserve the asymmetric cluster uncertainties. We then repeat the analysis with a stricter matching threshold, $\Delta D_C / D_C = 3\%$, and extend the dataset by including four DESI 2024 BAO measurements of $D_M/r_d$. For the BAO extension, we consider both a free-$r_d$ case and a case with a Planck prior on $r_d$.}
{Our analysis yields $\eta = 0.050^{+0.348}_{-0.307}$ and $\varepsilon = -0.184^{+0.724}_{-0.574}$ (68\% CL), showing no statistically significant evidence for CDDR violation or SNe~Ia evolution. Tightening the matching tolerance to $3\%$ leaves the conclusions unchanged. Including DESI BAO data does not qualitatively modify the inferred parameters. When $r_d$ is treated as a free parameter, we find $r_d = 140.56^{+8.12}_{-8.11}$, while imposing a Planck prior yields $r_d = 147.078\pm0.259$. In all cases, both $\eta = 0$ and $\varepsilon = 0$ remain consistent with the data. Additionally, we derive a cosmology-independent calibration of $M_0 = -19.460^{+0.126}_{-0.124}$ mag, consistent with local SH0ES measurements.}
{The standard cosmological framework remains robust under this model-independent scrutiny. Our results are stable against variations in the matching tolerance and the inclusion of current DESI BAO data. We further find that exploratory spline reconstructions introduce additional non-physical degeneracies in the present data regime, so that the matched-pair framework remains the most transparent and stable method for this analysis. By effectively breaking the degeneracy between opacity and evolution, our method offers a reliable pathway for future precision tests using larger samples from upcoming surveys.}


\maketitle

\section{Introduction}
\label{sec:intro}

The standard cosmological model, $\Lambda$CDM, has achieved remarkable success in explaining a wide range of observations, from the cosmic microwave background (CMB) anisotropies \cite{Planck2020} to the large-scale structure of the Universe \cite{Peebles1993}. However, with the advent of precision cosmology, increasing tensions---most notably the Hubble tension ($H_0$)---have emerged between early-universe predictions and late-universe measurements \cite{Riess2022, DiValentino2021, Abdalla2022}. Before attributing these discrepancies to exotic dark energy or modified gravity, it is imperative to scrutinize the fundamental observational principles that underpin our distance measurements \cite{Perivolaropoulos2022}. Among these, the cosmic distance-duality relation (CDDR) stands as a cornerstone of observational geometry \cite{Etherington1933}.

First derived by Etherington in 1933, the CDDR, also known as the reciprocity theorem, relates the luminosity distance $D_L$ and the angular diameter distance $D_A$ of a source at redshift $z$ via the simple identity:
\begin{equation}
	\frac{D_L}{D_A} (1+z)^{-2} = 1.
	\label{eq:cddr}
\end{equation}
The validity of this relation rests on three crucial theoretical pillars: (i) the spacetime is described by a Riemannian metric theory of gravity \cite{Ellis1971, Ellis2007}; (ii) photons propagate along unique null geodesics \cite{Uzan2004}; and (iii) the photon number is conserved during propagation \cite{Bassett2004}. Consequently, any observational violation of the CDDR would signal fundamental revisions to physics \cite{Barua2026}. Possible mechanisms include non-metric theories of gravity \cite{Hees2014, Vagnozzi2023}, the existence of gray dust in the intergalactic medium \cite{Corasaniti2006, Holanda2013}, or the coupling of photons with axion-like particles into which they might oscillate \cite{Tiwari2017, Keil2026}.

To test this relation, a phenomenological parameter $\eta$ is commonly introduced to quantify potential deviations, parameterized as $D_L = D_A (1+z)^{2+\eta}$ \cite{Holanda2010, Li2011}. Standard cosmology predicts $\eta = 0$ \cite{Ellis2007}. Observational constraints on $\eta$ require dual measurements of $D_L$ and $D_A$ at the same redshift \cite{Holanda2012}. While Type Ia Supernovae (SNe Ia) are the premier standard candles for $D_L$ \cite{Scolnic2022}, obtaining model-independent $D_A$ measurements remains challenging. Galaxy clusters (GCs) provide a powerful solution \cite{DeFilippis2005}. By combining X-ray surface brightness observations with the Sunyaev-Zel'dovich (SZ) effect---the inverse Compton scattering of CMB photons by hot intracluster gas \cite{Sunyaev1972}---the angular diameter distance to clusters can be determined directly, independent of the cosmic expansion history \cite{Cavaliere1976, Bonamente2006}.

Historically, tests of the CDDR have often relied on assuming a specific cosmological model (typically flat $\Lambda$CDM) to align datasets or to reconstruct distance functions \citep{Uzan2004, Holanda2010}. Within these model-dependent frameworks, deviations from the CDDR are frequently interpreted as signatures of cosmic opacity, a topic extensively explored in various cosmological contexts \citep[e.g.,][]{Hu2017}. Furthermore, such tests have been extended to alternative cosmological scenarios, including the $R_h=ct$ universe \citep{Hu2018}. However, these model-dependent approaches inevitably introduce circularity and bias if the assumed background cosmology deviates from reality \citep{Liang2013}. To circumvent this, model-independent reconstruction techniques, such as Gaussian Processes \citep{Seikel2012, Zhang2014, Keil2026}, polynomial fits \citep{Montiel2021}, or Padé approximants \citep{Barua2026}, have been widely adopted. Nevertheless, these methods can suffer from overfitting or boundary artifacts, particularly when data is sparse at high redshifts \citep{Kanodia2026}.

A more transparent approach is the ``matched pair'' technique pioneered by \citet{Holanda2010, Holanda2012}, where $D_L$ and $D_A$ measurements are strictly paired based on redshift proximity, thereby eliminating the dependence on the fiducial model entirely. To further improve the robustness of this technique, \citet{Zhou2021} introduced the dimensionless distance error consistency method, which allows for an adaptive matching tolerance based on data quality. This rigorous approach is arguably superior to simple redshift cuts and has been successfully applied in recent works \citep{Hu2023, Hu2023prd}. However, a common limitation in these previous analyses---including those utilizing the advanced matching criterion---is the neglect of the full systematic covariance matrix of SNe Ia during the analysis. In this work, we combine this advanced matching strategy with a rigorous treatment of the full Pantheon+ covariance matrix to ensure a stringent test.

A critical, yet often overlooked systematic in CDDR tests is the intrinsic evolution of the probes themselves \citep{Tutusaus2017}. While SNe Ia are standardized to a high degree, recent analyses suggest that their absolute magnitude ($M_B$) may evolve with redshift due to progenitor age or metallicity drift \citep{Kang2020, Nicolas2021, Scolnic2022}. Neglecting this evolution can mimic a violation of the CDDR, leading to spurious detections of $\eta \neq 0$ or biased constraints on opacity parameters \citep{Tutusaus2019, Kanodia2026}. Therefore, a rigorous test must break the degeneracy between the fundamental spacetime geometry (the CDDR parameter) and the astrophysical systematics (the source evolution) \citep{Alfano2026}.

In this work, we present a stringent, model-independent test of the CDDR using a carefully selected sample of 38 Galaxy Clusters providing $D_A$ measurements via the X-ray/SZ technique \citep{Bonamente2006}. We pair these clusters with the latest Type Ia Supernovae from the Pantheon+ compilation \citep{Scolnic2022}, which offers significantly reduced statistical and systematic uncertainties compared to previous samples. Unlike previous studies that often fix the supernova absolute magnitude, we adopt a joint analysis framework using Markov Chain Monte Carlo (MCMC) methods. We simultaneously constrain the CDDR parameter $\eta$ and the redshift-dependent evolution of the SNe Ia absolute magnitude, parameterized as $M_B(z) = M_0 + \varepsilon \cdot z$. By utilizing the covariance information of the Pantheon+ sample through the extracted sub-covariance matrix for the matched SNe~Ia, this study aims to provide robust, degeneracy-free constraints on the validity of the Etherington reciprocity theorem.

In addition to the baseline cluster--SNe~Ia matched-pair analysis, we further incorporate recent DESI BAO measurements in the form of $D_M/r_d$ \citep{DESI2024_III, DESI2024_VI}. This extension provides an additional geometric consistency check within the same framework, while preserving the model-independent nature of the test and allowing us to assess the robustness of the inferred constraints.

The structure of this paper is outlined as follows. Section \ref{sec:data} details the observational datasets utilized in this study, along with the specific methodology employed for data pairing. In Section \ref{sec:method}, we present the statistical framework and report the numerical constraints obtained from our analysis. Finally, Section \ref{sec:conclusions} summarizes our findings and discusses their implications.

\section{Data and Methodology}
\label{sec:data}
In this section, we provide a detailed description of the observational datasets employed in our analysis, including the galaxy cluster sample derived from X-ray and Sunyaev--Zel’dovich effect measurements and the Type Ia Supernova (SNe~Ia) compilation. We also introduce the DESI 2024 BAO measurements used as an extension of the baseline dataset. Furthermore, we outline the matching procedure used to construct pairs of angular diameter and luminosity distances at identical redshifts, which forms the basis of our model-independent test.

\subsection{Galaxy Cluster Sample}
\label{subsec:cluster_data}

For the angular diameter distance ($D_A$) measurements, we utilize the sample of 38 galaxy clusters compiled and analyzed by Bonamente et al. \cite{Bonamente2006}. This sample spans a redshift range of $0.14 \le z \le 0.89$ and represents one of the most robust datasets for SZE-based distance determinations.

The distances to these clusters were derived by combining X-ray surface brightness observations with Sunyaev-Zel'dovich effect (SZE) measurements. The X-ray data were obtained from the \textit{Chandra X-ray Observatory}, which provides high-resolution imaging and spectroscopy essential for characterizing the intracluster medium (ICM). The SZE data, which measure the spectral distortion of the Cosmic Microwave Background (CMB) caused by inverse Compton scattering off hot ICM electrons, were collected using the interferometric arrays at the Owens Valley Radio Observatory (OVRO) and the Berkeley-Illinois-Maryland Association (BIMA).

The fundamental principle of this distance measurement technique relies on the different dependencies of the SZE intensity and X-ray surface brightness on the electron density ($n_e$). The SZE temperature decrement is proportional to the line-of-sight integral of the electron pressure ($\Delta T_{\rm SZE} \propto \int n_e T_e \, dl$), whereas the X-ray surface brightness is proportional to the line-of-sight integral of the electron density squared ($S_X \propto \int n_e^2 \Lambda_{ee} \, dl$). By assuming a geometric model for the cluster (e.g., spherical symmetry) and modeling the radial profiles of the gas density and temperature, one can solve for the angular diameter distance $D_A$ directly, independent of the cosmic distance ladder or the expansion history of the Universe.

In the analysis by Bonamente et al. \cite{Bonamente2006}, a hydrostatic equilibrium model was employed to account for radial variations in plasma density, temperature, and metal abundance, thereby reducing systematic uncertainties associated with simpler isothermal $\beta$-models. The final sample consists of 38 clusters with high-quality joint X-ray and SZE detections, providing a reliable dataset for testing the cosmic distance-duality relation. It is worth noting that the derivation of $D_A$ from X-ray and SZE observations relies on standard local physics, including photon propagation and the thermodynamic properties of the intracluster medium, which are typically formulated within a metric theory of gravity such as General Relativity. However, this procedure does not rely on any assumption about the global expansion history, and therefore remains independent of any specific background cosmological model.

\subsection{Type Ia Supernova Sample}
\label{subsec:sn_data}

For the luminosity distance ($D_L$) measurements, we employ the latest \textit{Pantheon+} compilation, as presented by Brout et al. \cite{Brout2022} and Scolnic et al. \cite{Scolnic2022}. Pantheon+ represents a significant expansion and improvement over the original Pantheon sample, comprising 1701 light curves of 1550 distinct spectroscopically confirmed Type Ia supernovae (SNe Ia). The sample covers a broad redshift range of $0.001 < z < 2.26$, integrating data from multiple surveys including CfA1-4, CSP, PS1, SDSS, SNLS, and high-redshift samples from HST.

A key feature of Pantheon+ is its substantial increase in the number of low-redshift supernovae and its comprehensive treatment of systematic uncertainties. The compilation provides a detailed covariance matrix ($\mathbf{C}_{\rm stat+sys}$) that accounts for uncertainties arising from calibration, photometric zeropoints, intrinsic scatter, selection bias, and peculiar velocities. This robust quantification of errors is crucial for our joint analysis, as it allows us to propagate systematic correlations through the relevant sub-covariance matrix extracted from the full Pantheon+ covariance when constraining the CDDR parameter.

In the context of our matched-pair analysis, the Pantheon+ dataset is particularly advantageous because it includes the SH0ES (Supernovae and $H_0$ for the Equation of State of dark energy) subsample SNe Ia in galaxies that also host Cepheid variables. While the SH0ES calibration is typically used to fix the absolute magnitude $M_B$ for $H_0$ measurements, in this work, we treat the absolute magnitude and its potential redshift evolution ($\varepsilon$) as free nuisance parameters. We utilize the corrected apparent magnitudes ($m_B^{\rm corr}$) provided by the Pantheon+ analysis, which have been standardized for light-curve shape ($x_1$), color ($c$), and host-galaxy mass step corrections.

\subsection{Data Pairing}
\label{subsec:pairing}

To perform a model-independent test of the CDDR, we pair each galaxy cluster (providing $D_A$) with a Type Ia supernova (providing $D_L$) at (nearly) the same redshift. A key challenge in such analyses is the redshift mismatch between independent datasets, which can introduce non-negligible systematic biases if not properly controlled.

Instead of applying a simple redshift threshold (e.g., $\Delta z < 0.005$), we adopt a distance-based matching criterion, which more directly reflects the impact of redshift differences on cosmological distance measurements. This approach ensures that the pairing is performed in a physically meaningful way, minimizing the propagation of mismatch errors into the final constraints.

To implement this, we assume a fiducial flat $\Lambda$CDM cosmology with $\Omega_m = 0.3$ solely for the purpose of defining the matching tolerance. This assumption does not affect the model-independence of the final CDDR test, as it is only used to determine the relative distance differences between objects, not to derive the distances themselves. We compute the line-of-sight comoving distance, $D_C(z)$, for both the galaxy clusters and the supernovae, and require that a valid pair satisfies
\begin{equation}
	\frac{|D_C(z_{\rm cluster}) - D_C(z_{\rm SNe})|}{D_C(z_{\rm cluster})} \le 5\%,
	\label{eq:matching_criterion}
\end{equation}
which defines our baseline matching tolerance.

This dimensionless distance-based criterion has been shown to be more robust than simple redshift cuts, as it naturally accounts for the redshift dependence of cosmological distances. It has been successfully applied in previous studies, for example in constraining the mass density profiles of strong gravitational lensing systems \citep{Hu2023} and in model-independent calibrations of SNe Ia absolute magnitudes \citep{Hu2023prd}. 

To assess the robustness of our results against the choice of matching criterion, we additionally consider a stricter selection with a tolerance of $\Delta D_C / D_C \le 3\%$, and compare the resulting constraints with those obtained from the baseline sample.

From the pool of potential matches satisfying Eq.~(\ref{eq:matching_criterion}), we construct the final sample using a greedy selection algorithm. All candidate pairs are ranked according to their absolute distance difference $|D_C(z_{\rm cluster}) - D_C(z_{\rm SNe})|$, and the pair with the smallest difference is selected iteratively. Once a pair is selected, both the corresponding galaxy cluster and supernova are removed from the pool to ensure that each object is used at most once (sampling without replacement). This procedure minimizes the overall mismatch across the dataset and yields a well-defined set of independent pairs.

\begin{figure}[htbp]
	\centering
\includegraphics[width=0.5\textwidth]{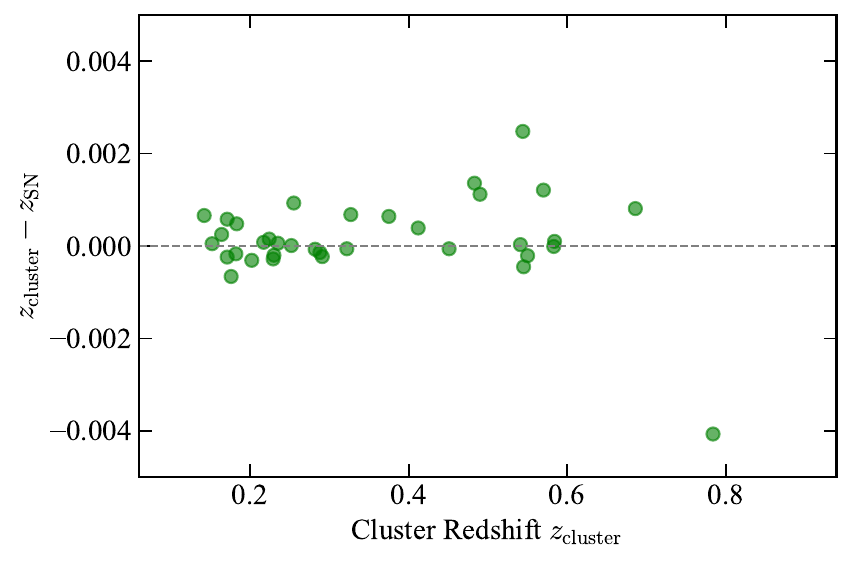}
\caption{Redshift mismatch ($\Delta z = z_{\rm cluster} - z_{\rm SN}$) between the paired galaxy clusters and SNe~Ia used in the baseline matched sample.}
\label{fig:matching}
\end{figure}

Using this method, we obtain a final matched sample of 38 galaxy cluster--SNe~Ia pairs, as illustrated in Fig.~\ref{fig:matching}. 

In parallel with the pair selection, we extract the corresponding $38 \times 38$ sub-covariance matrix from the full Pantheon+ covariance matrix ($\mathbf{C}_{\rm stat+sys}$). This step preserves both statistical and systematic correlations among the selected SNe~Ia, which is essential for a consistent likelihood analysis.

The final matched sample consists of 38 galaxy cluster--SNe~Ia pairs. The detailed properties of these pairs, including cluster redshifts, angular diameter distances, and the matched SNe~Ia corrected magnitudes, are listed in Table~\ref{tab:matched_data} in Appendix~\ref{app:data_table}.

\subsection{DESI BAO supplement}
\label{subsec:desi}

To address the request for additional angular-diameter-distance information, we supplemented the baseline cluster-only analysis with four DESI 2024 BAO measurements of $D_M/r_d$ at $z=0.510$, $0.706$, $0.930$, and $1.317$ \citep{DESI2024_III, DESI2024_VI}. These values were selected from the DESI Gaussian BAO summary products, and the corresponding $4\times4$ sub-covariance matrix was extracted from the full DESI covariance matrix. In our implementation this sub-covariance is diagonal for the selected set of $D_M/r_d$ points, but we nevertheless retain the formal matrix treatment for completeness. The corresponding BAO measurements are listed in Table~\ref{tab:desi}.

The BAO extension is used as a supplement to the matched-pair analysis rather than as a replacement for it. We consider two versions of this extension: one in which the sound horizon $r_d$ is treated as a free parameter, and one in which a Gaussian Planck prior is imposed on $r_d$. This allows us to separate the constraining power of the DESI points themselves from that of an external early-Universe calibration.

We emphasize that the BAO measurements do not directly constrain the CDDR parameter $\eta$, but instead provide additional geometric information that helps to break degeneracies among the model parameters. In this sense, the DESI BAO data act as a complementary consistency check within the same distance framework, rather than an independent probe of the CDDR.

\begin{table}
	\caption{\textbf{DESI BAO points used in the supplementary analysis.}}
	\label{tab:desi}
	\centering
	\begin{tabular}{ccc}
		\hline\hline
		$z$ & Observable & Value \\
		\hline
		0.510 & $D_M/r_d$ & 13.6200 \\
		0.706 & $D_M/r_d$ & 16.8465 \\
		0.930 & $D_M/r_d$ & 21.7084 \\
		1.317 & $D_M/r_d$ & 27.7872 \\
		\hline
	\end{tabular}
\end{table}

\section{Analysis and Results}
\label{sec:method}

\subsection{Statistical Method}
\label{subsec:stat_method}

We adopt a joint likelihood analysis to simultaneously constrain the CDDR parameter $\eta$, the SNe~Ia absolute magnitude $M_B$, and its evolutionary parameter $\varepsilon$.
The distance duality relation is parameterized as:
\begin{equation}
	D_L(z) = D_A(z) (1+z)^2 (1+\eta z).
	\label{eq:cddr_new}
\end{equation}
Ideally, if the reciprocity theorem holds, $\eta = 0$.

For the Type Ia supernovae, the observed distance modulus including redshift evolution is written as:
\begin{equation}
	\mu_{\rm SN}^{\rm obs}(z_i) = m_B^{\rm corr} - (M_0 + \varepsilon z_i),
\end{equation}
where $M_0$ is the absolute magnitude at $z=0$, and $\varepsilon$ quantifies a possible linear evolution.

For the galaxy clusters, the observed angular diameter distance $D_A$ is converted to a distance modulus prediction using the modified CDDR parameter $\eta$:
\begin{equation}
	\mu_{\rm cluster}^{\rm th}(z_i, \eta) = 5 \log_{10} \left[ D_A^{\rm obs}(z_i) (1+z_i)^2 (1+\eta z_i) \right] + 25.
\end{equation}

The galaxy cluster angular diameter distance measurements possess asymmetric statistical uncertainties ($\sigma_{D_A,+}$ and $\sigma_{D_A,-}$). To rigorously incorporate this non-Gaussian feature, we utilize a split-normal (or piecewise Gaussian) likelihood formulation. First, we propagate these asymmetric errors from linear distance space ($D_A$) to distance modulus space ($\mu$) using error propagation: $\sigma_{\mu,\pm} \approx (5/\ln 10) (\sigma_{D_A,\pm} / D_A)$.

In our joint analysis, the total covariance matrix $\mathbf{C}_{\rm tot}$ combines the supernova sub-covariance matrix (extracted from the full Pantheon+ covariance) and the cluster diagonal variances.

The asymmetric uncertainties in clusters arise from the non-linear propagation of systematic errors in X-ray temperature and SZE measurements \citep{Bonamente2006}. To strictly preserve this information, the variance term for the $i$-th cluster, $\sigma_{i,\rm cluster}^2$, is selected dynamically based on the residual direction:
\begin{equation}
	\sigma_{i,\rm cluster} = 
	\begin{cases} 
		\sigma_{\mu,i,+} & \text{if } \mu_{\rm SN}^{\rm obs}(z_i) > \mu_{\rm cluster}^{\rm th}(z_i), \\
		\sigma_{\mu,i,-} & \text{otherwise}.
	\end{cases}
	\label{eq:split_normal}
\end{equation}
This formulation ensures that the likelihood properly accounts for the asymmetric error distribution of the cluster measurements.

We define the residuals vector as
\begin{equation}
	\boldsymbol{\Delta} = \boldsymbol{\mu}_{\rm SN}^{\rm obs} - \boldsymbol{\mu}_{\rm cluster}^{\rm th}.
\end{equation}

We assume that the observational uncertainties of the galaxy clusters and the SNe~Ia are independent, since they arise from different instruments and physical processes (X-ray/SZ observations versus optical light curves). Therefore, no cross-covariance term between cluster and SN measurements is included.

The final log-likelihood function is constructed as:
\begin{equation}
	\ln \mathcal{L} = -\frac{1}{2} \left[ \boldsymbol{\Delta}^T \mathbf{C}_{\rm tot}^{-1} \boldsymbol{\Delta} + \ln(\det \mathbf{C}_{\rm tot}) + N \ln(2\pi) \right].
\end{equation}

We perform the parameter estimation using the Markov Chain Monte Carlo (MCMC) method with the \texttt{emcee} Python package. We assume flat priors for all parameters: $M_0 \in [-20.5, -18.0]$, $\varepsilon \in [-1.0, 1.0]$, and $\eta \in [-2.0, 2.0]$.

\medskip

For the DESI-extended analysis, the parameter set becomes $\{M_0, \varepsilon, \eta, r_d\}$. We verified that the 38 SNe~Ia matched to the galaxy clusters are distinct from the 4 SNe~Ia used for the BAO anchor, ensuring no object-level overlap.

Although both subsets originate from the Pantheon+ compilation, a residual cross-covariance may arise from shared calibration systematics. However, given the small number of BAO-anchoring SNe~Ia and the relatively large uncertainties of the cluster measurements, this contribution is subdominant and does not affect the parameter inference at the current level of precision. Therefore, we neglect this cross-covariance and treat $\chi^2_{\rm cl}$ and $\chi^2_{\rm BAO}$ as additive.

In our implementation, the theoretical prediction for the BAO observable is constructed from the SN-inferred luminosity distance combined with the deformed CDDR relation.

The SNe~Ia distance modulus at the DESI redshifts is written as
\begin{equation}
	\mu_{\rm SN}(z) = m_B - (M_0 + \varepsilon z),
\end{equation}

from which the luminosity distance is obtained as
\begin{equation}
	D_L(z) = 10^{(\mu_{\rm SN} - 25)/5}.
\end{equation}

Using the generalized CDDR relation, the transverse comoving distance is
\begin{equation}
	D_M(z) = \frac{D_L(z)}{(1+z)(1+\eta z)}.
\end{equation}

The theoretical BAO observable is therefore
\begin{equation}
	\left( \frac{D_M}{r_d} \right)_{\rm th}
	=
	\frac{1}{r_d}\,
	\frac{D_L(z)}{(1+z)(1+\eta z)}.
\end{equation}

The residual vector is defined as
\begin{equation}
	\Delta_{\rm BAO} =
	\left( \frac{D_M}{r_d} \right)_{\rm obs}
	-
	\left( \frac{D_M}{r_d} \right)_{\rm th}.
\end{equation}

The total covariance matrix is constructed as
\begin{equation}
	C_{\rm BAO} = C_{\rm DESI} + J C_{\rm SN} J^T,
\end{equation}

where $C_{\rm DESI} = \mathrm{diag}(\sigma_i^2)$ and the Jacobian matrix $J$ propagates the SN covariance into the BAO observable space,
\begin{equation}
	J_{ii} = \frac{\ln 10}{5} \left( \frac{D_M}{r_d} \right)_i.
\end{equation}

The BAO log-likelihood is written as
\begin{equation}
	\ln \mathcal{L}_{\rm BAO} =
	-\frac{1}{2}
	\left[
	\Delta_{\rm BAO}^T C_{\rm BAO}^{-1} \Delta_{\rm BAO}
	+ \ln \det C_{\rm BAO}
	+ N \ln (2\pi)
	\right].
\end{equation}

We consider both the free-$r_d$ case and the case with a Planck prior:
\begin{equation}
	\chi^2_{\rm prior} = \left( \frac{r_d - 147.09}{0.26} \right)^2.
\end{equation}

\subsection{Baseline matched-pair constraints}
\label{subsec:baseline_results}

The constraints for the baseline 5\% matched sample are:
\begin{align}
M_0 &= -19.460^{+0.126}_{-0.124}, \nonumber\\
\varepsilon &= -0.184^{+0.724}_{-0.574}, \nonumber\\
\eta &= 0.050^{+0.348}_{-0.307}.
\end{align}
The best-fit log-likelihood is $\ln\mathcal{L}_{\rm best}=-44.56$, with $\mathrm{AIC}=95.12$ and $\mathrm{BIC}=100.03$.

\begin{figure}[htbp]
	\centering
\includegraphics[width=0.5\textwidth]{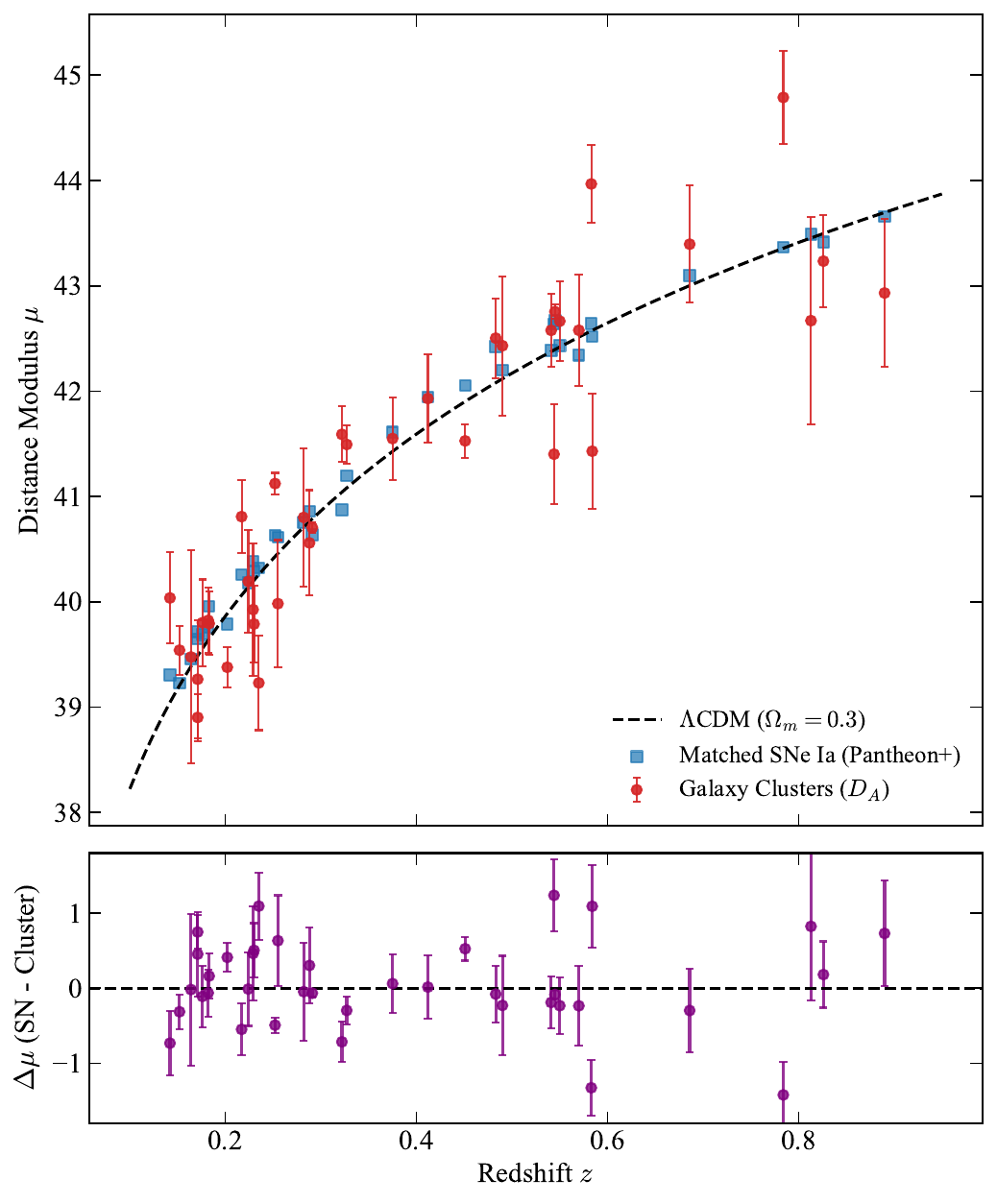}
	\caption{The results of the joint analysis. Upper panel: The Hubble diagram for the 38 matched pairs. The red circles represent the distance moduli of galaxy clusters (derived from $D_A$ assuming $\eta = 0$), and the blue squares represent the matched Pantheon+ SNe Ia. The dashed line indicates the prediction of the standard flat $\Lambda$CDM model ($\Omega_m = 0.3$) for reference. Lower panel: The distance-modulus residuals ($\Delta \mu = \mu_{\rm SNe} - \mu_{\rm Cluster}$) as a function of redshift. The data points scatter around zero with no significant redshift dependence, supporting the validity of the CDDR and the absence of strong SNe~Ia evolution. Error bars represent the asymmetric uncertainties of the cluster distance moduli. The Pantheon+ SNe Ia uncertainties are not shown individually for clarity, but are fully accounted for in the covariance matrix used in the likelihood analysis.}
\label{fig:hubble_resid} 
\end{figure}

\begin{figure}[htbp]
	\centering
	\includegraphics[width=0.5\textwidth]{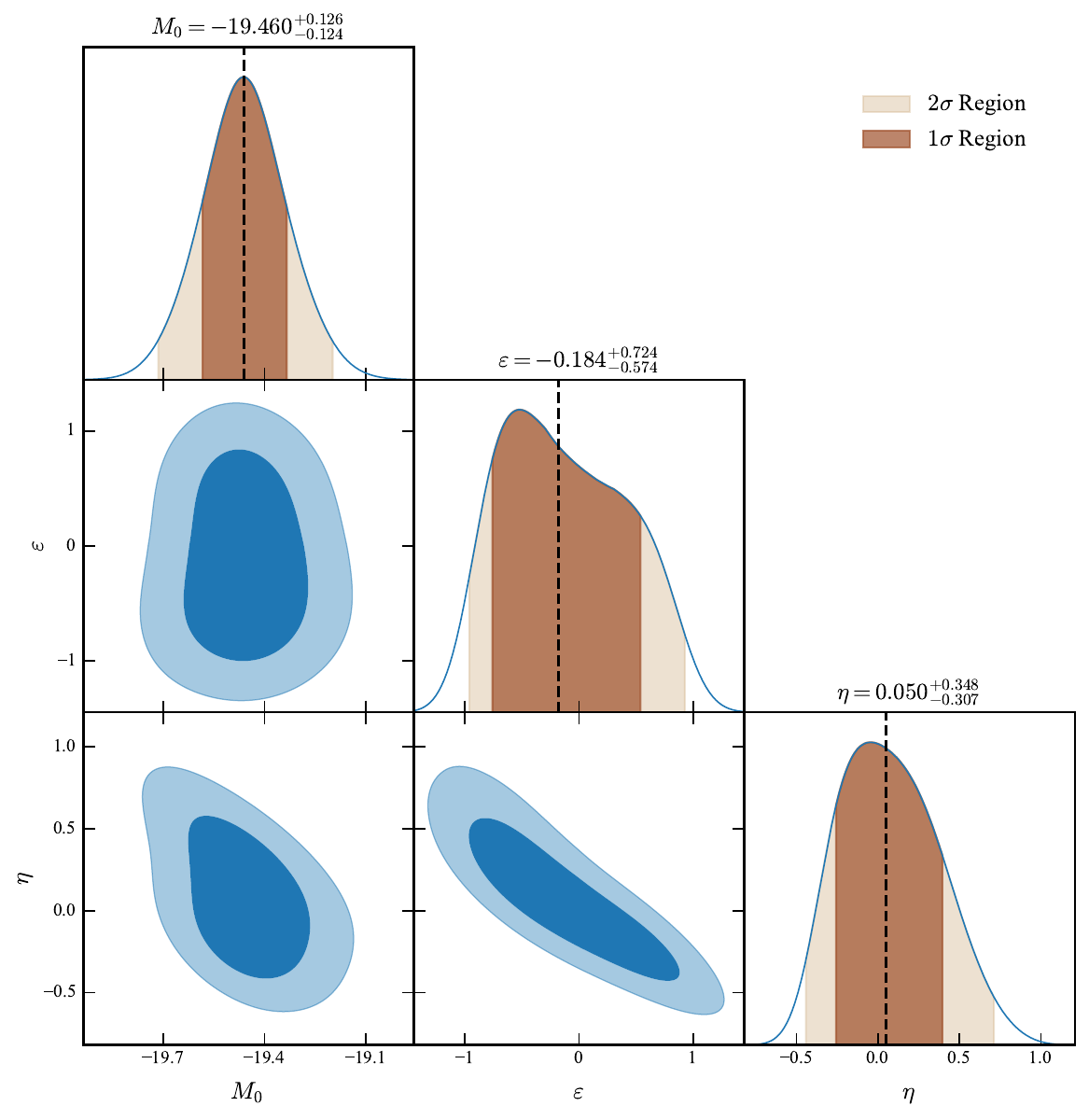}
\caption{The one-dimensional and two-dimensional marginalized posterior distributions for the parameters $M_0$, $\varepsilon$, and $\eta$. The contours represent the 68\% (1$\sigma$) and 95\% (2$\sigma$) confidence levels, respectively. The diagonal panels show the marginalized probability density functions with vertical dashed lines indicating the median values. A slight degeneracy is observed between the evolution parameter $\varepsilon$ and the CDDR-violation parameter $\eta$, but the results remain consistent with the standard CDDR ($\eta = 0$) within 1$\sigma$, indicating no evidence for CDDR violation.}
\label{fig:contours}
\end{figure}

These results are derived from the 38 matched galaxy cluster--SNe~Ia pairs spanning the redshift range $0.14 \lesssim z \lesssim 0.89$. These baseline constraints already show that neither a CDDR violation nor a significant redshift evolution of the standardized SNe~Ia absolute magnitude is required by the data. The posterior asymmetry visible in the one-dimensional distributions arises naturally from the split-normal treatment of the cluster uncertainties and is therefore a feature of the data model rather than an MCMC artifact. The corresponding constraints are summarized in Table~\ref{tab:tolerance} and illustrated in Fig.~\ref{fig:contours}. The quality of the fit is further illustrated by the Hubble diagram shown in Fig.~\ref{fig:hubble_resid}.

\subsection{Robustness against the matching tolerance}
\label{subsec:tolerance_results}

\begin{figure}[htbp]
	\centering
	\includegraphics[width=0.5\textwidth]{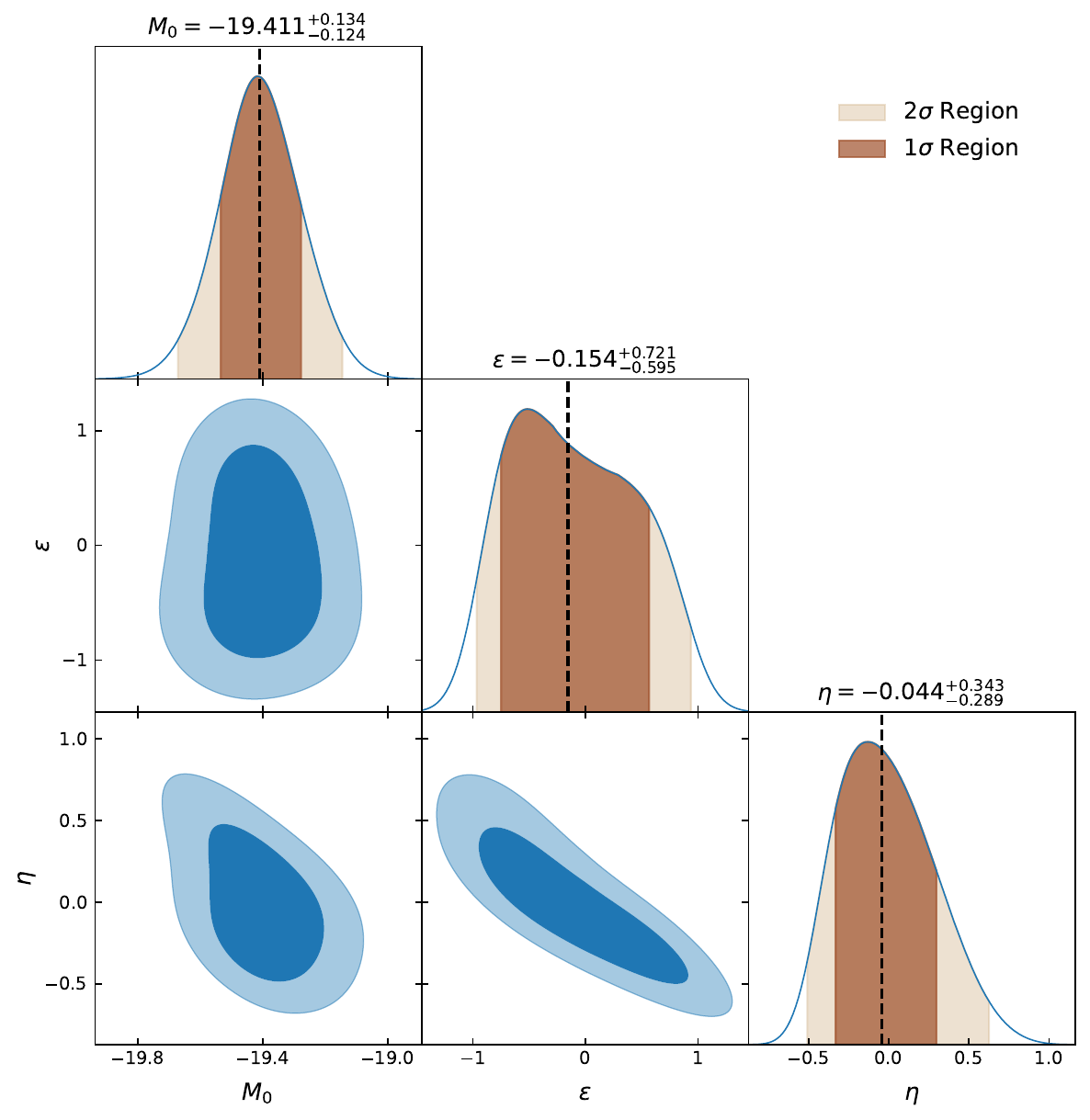}
\caption{Posterior constraints for the stricter $\Delta D_C / D_C \leq 3\%$ matched sample, showing consistency with the baseline 5\% results.}
\label{fig:tolerance03}
\end{figure}

To test the robustness of our results against the choice of matching threshold, we repeated the full cluster-only matched-pair analysis with a stricter tolerance of $\Delta D_C / D_C = 3\%$. In this case, 37 clusters remain in the sample, and the resulting constraints are
\begin{align}
M_0 &= -19.411^{+0.134}_{-0.124}, \nonumber\\
\varepsilon &= -0.154^{+0.721}_{-0.595}, \nonumber\\
\eta &= -0.044^{+0.343}_{-0.289}.
\end{align}
The best-fit likelihood is $\ln\mathcal{L}_{\rm best}=-43.01$, with $\mathrm{AIC}=92.03$ and $\mathrm{BIC}=96.86$. This test directly probes the sensitivity of the matched-pair method to the redshift-matching criterion. These results are statistically indistinguishable from the baseline $5\%$ case. These results are statistically indistinguishable from the baseline case. A comparison between the $5\%$ and $3\%$ matching cases is presented in Table~\ref{tab:tolerance}. Since all 38 clusters are retained for any tolerance $\ge5\%$, the corresponding matched dataset and inferred constraints are identical to the baseline case. This shows that the conclusions are not sensitive to the exact choice of matching tolerance once the threshold is at or above $5\%$. This stability is also clearly illustrated in Fig.~\ref{fig:contours} and Fig.~\ref{fig:tolerance03}.

\begin{table*}[htbp]
\centering
\caption{Robustness of the cluster-only matched-pair analysis against the matching tolerance.}
\label{tab:tolerance}
\begin{tabular}{lcccccc}
\hline\hline
Tolerance & Matched clusters & $M_0$ & $\varepsilon$ & $\eta$ & AIC & BIC \\
\hline
$3\%$ & $37$ & $-19.411^{+0.134}_{-0.124}$ & $-0.154^{+0.721}_{-0.595}$ & $-0.044^{+0.343}_{-0.289}$ & $92.03$ & $96.86$ \\
$5\%$ & $38$ & $-19.460^{+0.126}_{-0.124}$ & $-0.184^{+0.724}_{-0.574}$ & $0.050^{+0.348}_{-0.307}$ & $95.12$ & $100.03$ \\
$\ge5\%$ & $38$ & same as $5\%$ & same as $5\%$ & same as $5\%$ & same as $5\%$ & same as $5\%$ \\
\hline\hline
\end{tabular}
\end{table*}

\subsection{Supplementary constraints from DESI BAO}
\label{subsec:desi_results}

\begin{figure}[htbp]
	\centering
	\includegraphics[width=0.5\textwidth]{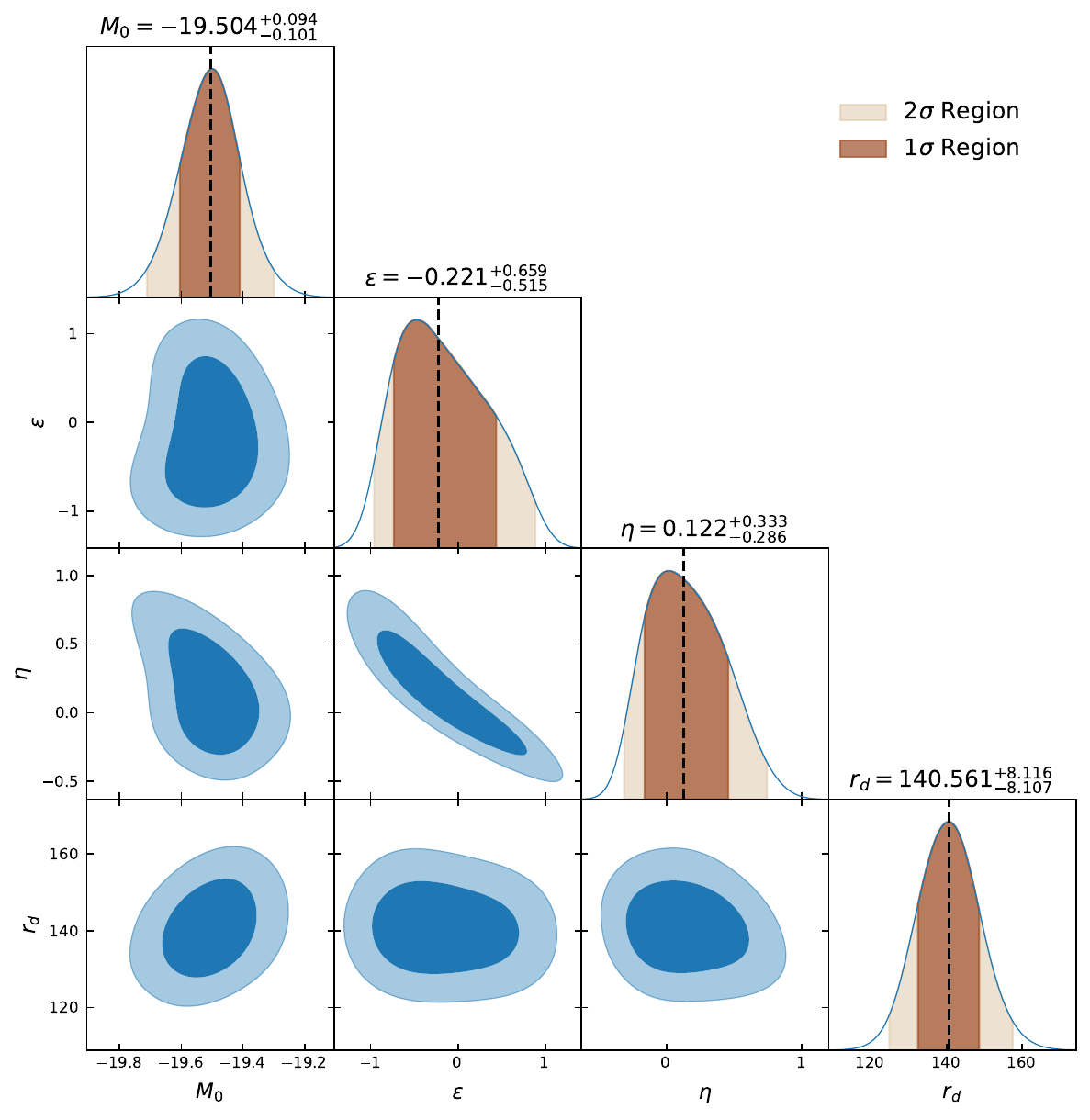}
\caption{Posterior constraints after including DESI BAO data with the sound horizon $r_d$ treated as a free parameter (prior-free case), demonstrating that the inferred parameters remain consistent with the cluster-only analysis.}
\label{fig:desi_post}
\end{figure}

We next augment the matched-pair analysis with four DESI 2024 BAO measurements. When the sound horizon is left completely free, we obtain
\begin{equation}
	\begin{aligned}
		M_0 &= -19.504^{+0.094}_{-0.101},\\
		\varepsilon &= -0.221^{+0.659}_{-0.515},\\
		\eta &= 0.123^{+0.333}_{-0.286},\\
		r_d &= 140.56^{+8.12}_{-8.11}\,\mathrm{Mpc}.
	\end{aligned}
\end{equation}
This result shows that the four selected BAO points provide some additional geometric information, but do not by themselves tightly constrain $r_d$. This is expected given the limited number of BAO points and the fact that the observable $D_M/r_d$ primarily constrains the overall distance scale rather than the CDDR deformation directly. Most importantly, the inferred values of $\eta$ and $\varepsilon$ remain fully consistent with the baseline matched-pair analysis.If instead we impose a Gaussian Planck prior on $r_d$, the posterior becomes
\begin{equation}
	\begin{aligned}
		M_0 &= -19.481^{+0.086}_{-0.093},\\
		\varepsilon &= -0.239^{+0.645}_{-0.519},\\
		\eta &= 0.079^{+0.309}_{-0.267},\\
		r_d &= 147.078 \pm 0.259\,\mathrm{Mpc}.
	\end{aligned}
\end{equation}
The corresponding best-fit values are $\ln\mathcal{L}_{\rm best}=-51.48$, $\mathrm{AIC}=110.95$, and $\mathrm{BIC}=117.90$ for the free-$r_d$ case, and $\ln\mathcal{L}_{\rm best}=-51.83$, $\mathrm{AIC}=111.66$, and $\mathrm{BIC}=118.61$ for the Planck-prior case. The combined constraints from the baseline, tolerance, and DESI-extended analyses are summarized in Table~\ref{tab:main_results}. We also repeated the DESI+Planck analysis for the stricter $3\%$ matched sample, obtaining $M_0=-19.472^{+0.086}_{-0.093}$, $\varepsilon=-0.249^{+0.639}_{-0.506}$, $\eta=0.069^{+0.292}_{-0.265}$, and $r_d=147.084\pm0.259$, again confirming the stability of the conclusions. The consistency of the results before and after including DESI BAO data is illustrated in Fig.~\ref{fig:desi_post}.

\begin{table*}[htbp]
\centering
\caption{Summary of the main constraints obtained in this work.}
\label{tab:main_results}
\begin{tabular}{lccccc}
\hline\hline
Dataset / method & $M_0$ & $\varepsilon$ & $\eta$ & $r_d$ & Comments \\
\hline
Clusters only, $5\%$ & $-19.460^{+0.126}_{-0.124}$ & $-0.184^{+0.724}_{-0.574}$ & $0.050^{+0.348}_{-0.307}$ & -- & baseline matched sample \\
Clusters only, $3\%$ & $-19.411^{+0.134}_{-0.124}$ & $-0.154^{+0.721}_{-0.595}$ & $-0.044^{+0.343}_{-0.289}$ & -- & robustness test \\
Clusters + DESI, free $r_d$ & $-19.504^{+0.094}_{-0.101}$ & $-0.221^{+0.659}_{-0.515}$ & $0.123^{+0.333}_{-0.286}$ & $140.56^{+8.12}_{-8.11}$ & supplementary BAO extension \\
Clusters + DESI + Planck & $-19.481^{+0.086}_{-0.093}$ & $-0.239^{+0.645}_{-0.519}$ & $0.079^{+0.309}_{-0.267}$ & $147.078\pm0.259$ & supplementary BAO extension \\
Clusters($3\%$) + DESI + Planck & $-19.472^{+0.086}_{-0.093}$ & $-0.249^{+0.639}_{-0.506}$ & $0.069^{+0.292}_{-0.265}$ & $147.084\pm0.259$ & joint robustness test \\
\hline\hline
\end{tabular}
\end{table*}

\subsection{Comment on spline reconstruction}
\label{subsec:spline}

To assess alternative, continuous distance representations, we explored a cubic-spline reconstruction using the full Pantheon+ sample. Our motivation was to test whether the matched-pair selection itself could be bypassed in favor of a continuous reconstruction. In practice, however, we found that the spline-based results depend sensitively on the adopted parameterization. One implementation led to an apparently tight but strongly shifted posterior for $\varepsilon$, while an alternative parameterization produced pathological drifts in the absolute scale. This indicates that, for the present dataset, the spline nodes compete with $M_0$ and $\varepsilon$ in a way that introduces additional non-physical degeneracies.

Because our scientific goal is to test the CDDR while preserving a transparent interpretation of the nuisance parameters, we do not adopt the spline reconstruction as part of the baseline inference. Instead, we regard it as an exploratory exercise whose main lesson is methodological: with the current sample size and error budget, the matched-pair approach is more stable and easier to interpret physically.

\section{Conclusions and Discussions}
\label{sec:conclusions}

The original motivation of this work was to test the CDDR while simultaneously marginalizing over a possible redshift evolution of the standardized SNe~Ia absolute magnitude. This remains one of the key physical aspects of the analysis: the parameters $\eta$ and $\varepsilon$ are not independent in practice, and neglecting one of them can bias the inference on the other.

A key aspect of our analysis is the intrinsic degeneracy between the CDDR-violation parameter $\eta$ and the SNe~Ia evolution parameter $\varepsilon$. This degeneracy is physically well motivated: a violation of photon conservation (e.g., negative $\eta$) can be observationally mimicked by an apparent brightening of SNe~Ia at high redshift (positive $\varepsilon$). Previous studies often fixed $\varepsilon=0$, leading to apparently tighter constraints on $\eta$. However, such assumptions risk introducing biases if source evolution is present. By jointly fitting $\eta$ and $\varepsilon$, our constraints are necessarily broader but significantly more robust against astrophysical systematics.

Our analysis can be summarized in three aspects. First, we explicitly clarify that the cluster-based distances are independent of the assumed background cosmology, but still depend on astrophysical assumptions about cluster structure and hydrostatic equilibrium. Second, we demonstrate that the matched-pair result is robust against the choice of matching tolerance: tightening the threshold from $5\%$ to $3\%$ changes the sample from 38 to 37 pairs, but leaves the inferred posterior for $\eta$ statistically unchanged. Third, we supplement the cluster-only analysis with four DESI BAO measurements and show that the central conclusion is unchanged both when $r_d$ is kept free and when it is anchored by a Planck prior.

Our main conclusions can be summarized as follows:
\begin{enumerate}
	\item There is no statistically significant evidence for a violation of the CDDR. In all matched-pair and DESI-augmented analyses, the parameter $\eta$ remains consistent with zero at the $1\sigma$ level.
	\item There is no significant evidence for redshift evolution of the standardized SNe~Ia absolute magnitude. The parameter $\varepsilon$ remains consistent with zero in all matched-pair analyses.
	\item The conclusions are robust against the matching tolerance. The stricter $3\%$ threshold produces constraints fully consistent with the baseline $5\%$ case.
	\item The inclusion of the present DESI BAO information modestly improves the geometric leverage of the analysis, but does not qualitatively alter the inferred values of $\eta$ and $\varepsilon$.
	\item Exploratory spline reconstructions are highly parameterization-dependent in the current data regime and introduce additional non-physical degeneracies. For this reason, the matched-pair method remains the most transparent and stable baseline method for the present study.
\end{enumerate}

An additional by-product of the analysis is the calibration of $M_0$ in a framework that does not rely on the local distance ladder. In the baseline matched-pair analysis we obtain $M_0=-19.460^{+0.126}_{-0.124}$, and the extended analyses give consistent values. This provides an independent consistency check on the standard SNe~Ia calibration, although current cluster uncertainties remain too large for a precision determination competitive with the local distance ladder. We emphasize that all tested extensions, including stricter matching criteria and the inclusion of DESI BAO data, lead to statistically consistent results, reinforcing the robustness of the matched-pair approach.

Future progress will come primarily from larger and cleaner cluster samples. Surveys such as eROSITA are expected to expand the number of well-characterized clusters dramatically \citep{Merloni2012}, while LSST and other time-domain surveys will vastly increase the number of SNe~Ia available for matched or near-matched analyses. In that regime it may become worthwhile to revisit reconstruction-based methods. For the present data quality, however, the matched-pair framework offers the best compromise between transparency, robustness, and physical interpretability.

\begin{acknowledgements}
We acknowledge the use of the Pantheon+ supernova data and the galaxy cluster compilation. This work is supported by Yunnan Youth Basic Research Projects 202501AT070439, the National Natural Science Foundation of China (No.12473029), Dali Expert Workstation of Rainer Spurzem, Yunnan Academician Workstation of Wang Jingxiu (202005AF150025), China Manned Space Project (No. CMS-CSST-2021-A08), Guanghe Foundation (No. ghfund202407013470), Jiangsu Funding Program for Excellent Postdoctoral Talent (20220ZB59), and China Postdoctoral Science Foundation (2022M721561).
\end{acknowledgements}

\bibliographystyle{aa}
\bibliography{references}

\par
\vspace{1cm}
\setcounter{section}{0}
\renewcommand{\thesection}{\Alph{section}}

\section{Matched Galaxy Cluster and SNe Ia Sample}
\label{app:data_table}

In this appendix, we present the catalog of the 38 matched pairs used in our baseline analysis. For each galaxy cluster, we list its redshift ($z_{\rm cl}$), angular diameter distance ($D_A$) with asymmetric uncertainties, and the derived distance modulus ($\mu_{\rm cl}$). The matched SNe~Ia properties, including redshift ($z_{\rm SN}$), SNe~Ia name (CID), and corrected apparent magnitude ($m_B$), are also provided.

\begin{table*}[h!]
\caption{The matched sample of 38 Galaxy Clusters and Type Ia Supernovae.}
\label{tab:matched_data}
\centering
\begin{tabular}{lcccccc}
\hline \hline
Cluster Name & $z_{\rm cl}$ & $D_A$ (Gpc) & $z_{\rm SN}$ & SN Name & $m_B$ (mag) & $\mu_{\rm cl}$ (mag) \\
\hline
Abell 1413  & 0.142 & $0.78^{+0.18}_{-0.13}$ & 0.1413 & 1317666 & 19.871 & $40.04^{+0.50}_{-0.36}$ \\
Abell 2204  & 0.152 & $0.61^{+0.06}_{-0.07}$ & 0.1520 & 2031 & 19.795 & $39.54^{+0.21}_{-0.25}$ \\
Abell 2259  & 0.164 & $0.58^{+0.29}_{-0.25}$ & 0.1638 & 19174 & 20.028 & $39.48^{+1.09}_{-0.94}$ \\
Abell 1914  & 0.171 & $0.44^{+0.04}_{-0.05}$ & 0.1704 & 12856 & 20.221 & $38.90^{+0.20}_{-0.25}$ \\
Abell 586  & 0.171 & $0.52^{+0.15}_{-0.12}$ & 0.1712 & 590005 & 20.289 & $39.27^{+0.63}_{-0.50}$ \\
Abell 2218  & 0.176 & $0.66^{+0.14}_{-0.11}$ & 0.1767 & 14979 & 20.263 & $39.80^{+0.46}_{-0.36}$ \\
Abell 665  & 0.182 & $0.66^{+0.09}_{-0.10}$ & 0.1822 & 17332 & 20.333 & $39.82^{+0.30}_{-0.33}$ \\
Abell 1689  & 0.183 & $0.65^{+0.09}_{-0.09}$ & 0.1825 & 15453 & 20.529 & $39.79^{+0.30}_{-0.30}$ \\
Abell 2163  & 0.202 & $0.52^{+0.04}_{-0.05}$ & 0.2023 & 13152 & 20.365 & $39.38^{+0.17}_{-0.21}$ \\
Abell 773  & 0.217 & $0.98^{+0.17}_{-0.14}$ & 0.2169 & 7473 & 20.840 & $40.81^{+0.38}_{-0.31}$ \\
Abell 2261  & 0.224 & $0.73^{+0.20}_{-0.13}$ & 0.2238 & 330083 & 20.763 & $40.19^{+0.59}_{-0.39}$ \\
Abell 2111 & 0.229 & $0.64^{+0.20}_{-0.17}$ & 0.2293 & 13072 & 20.970 & $39.93^{+0.68}_{-0.58}$ \\
Abell 267  & 0.230 & $0.60^{+0.11}_{-0.09}$ & 0.2302 & 21033 & 20.875 & $39.79^{+0.40}_{-0.33}$ \\
RX J2129.7+0005 & 0.235 & $0.46^{+0.11}_{-0.08}$ & 0.2349 & 1319366 & 20.905 & $39.23^{+0.52}_{-0.38}$ \\
Abell 1835 & 0.252 & $1.07^{+0.02}_{-0.08}$ & 0.2520 & 17552 & 21.217 & $41.12^{+0.04}_{-0.16}$ \\
Abell 68  & 0.255 & $0.63^{+0.16}_{-0.19}$ & 0.2541 & 5736 & 21.202 & $39.98^{+0.55}_{-0.65}$ \\
Abell 697 & 0.282 & $0.88^{+0.30}_{-0.23}$ & 0.2821 & 1317286 & 21.346 & $40.80^{+0.74}_{-0.57}$ \\
Abell 611  & 0.288 & $0.78^{+0.18}_{-0.18}$ & 0.2881 & 1334644 & 21.454 & $40.56^{+0.50}_{-0.50}$ \\
ZW 3146 & 0.291 & $0.83^{+0.02}_{-0.02}$ & 0.2912 & 19027 & 21.231 & $40.70^{+0.05}_{-0.05}$ \\
Abell 1995  & 0.322 & $1.19^{+0.15}_{-0.14}$ & 0.3221 & 120085 & 21.474 & $41.59^{+0.27}_{-0.26}$ \\
MS 1358.4+6245  & 0.327 & $1.13^{+0.09}_{-0.10}$ & 0.3263 & 1308326 & 21.797 & $41.49^{+0.17}_{-0.19}$ \\
Abell 370  & 0.375 & $1.08^{+0.19}_{-0.20}$ & 0.3744 & 16232 & 22.219 & $41.55^{+0.38}_{-0.40}$ \\
MACS J2228.5+2036 & 0.412 & $1.22^{+0.24}_{-0.23}$ & 0.4116 & 520071 & 22.560 & $41.93^{+0.43}_{-0.41}$ \\
RX J1347.5-1145  & 0.451 & $0.96^{+0.06}_{-0.08}$ & 0.4511 & 520016 & 22.675 & $41.53^{+0.14}_{-0.18}$ \\
MACS J2214.9-1359 & 0.483 & $1.44^{+0.27}_{-0.23}$ & 0.4816 & 120044 & 23.052 & $42.50^{+0.41}_{-0.35}$ \\
MACS J1311.0-0310 & 0.490 & $1.38^{+0.47}_{-0.37}$ & 0.4889 & 05D1ix & 22.831 & $42.43^{+0.74}_{-0.58}$ \\
CL 0016+1609 & 0.541 & $1.38^{+0.22}_{-0.22}$ & 0.5410 & 110426 & 23.026 & $42.58^{+0.35}_{-0.35}$ \\
MACS J1149.5+2223  & 0.544 & $0.80^{+0.19}_{-0.16}$ & 0.5415 & 1324542 & 23.277 & $41.40^{+0.52}_{-0.43}$ \\
MACS J1423.8+2404 & 0.545 & $1.49^{+0.06}_{-0.03}$ & 0.5454 & 520041 & 23.310 & $42.76^{+0.09}_{-0.04}$ \\
MS 0451.6-0305  & 0.550 & $1.42^{+0.26}_{-0.23}$ & 0.5502 & 050201 & 23.072 & $42.66^{+0.40}_{-0.35}$ \\
MACS J2129.4-0741  & 0.570 & $1.33^{+0.37}_{-0.28}$ & 0.5688 & 1317454 & 22.988 & $42.58^{+0.60}_{-0.46}$ \\
MS 2053.7-0449 & 0.583 & $2.48^{+0.41}_{-0.44}$ & 0.5830 & 04D1jg & 23.288 & $43.97^{+0.36}_{-0.39}$ \\
MACS J0647.7+7015 & 0.584 & $0.77^{+0.21}_{-0.18}$ & 0.5839 & 1338387 & 23.167 & $41.43^{+0.59}_{-0.51}$ \\
MACS J0744.8+3927 & 0.686 & $1.68^{+0.48}_{-0.38}$ & 0.6852 & 04D4ic & 23.764 & $43.40^{+0.62}_{-0.49}$ \\
MS 1137.5+6625 & 0.784 & $2.85^{+0.52}_{-0.63}$ & 0.7881 & 05D4cs & 24.051 & $44.79^{+0.40}_{-0.48}$ \\
RX J1716.4+6708 & 0.813 & $1.04^{+0.51}_{-0.43}$ & 0.7986 & 03D1fq & 24.181 & $42.67^{+1.06}_{-0.90}$ \\
MS 1054.5-0321  & 0.826 & $1.33^{+0.28}_{-0.26}$ & 0.8398 & Elvis & 24.107 & $43.23^{+0.46}_{-0.42}$ \\
CL J1226.9+3332  & 0.890 & $1.08^{+0.42}_{-0.28}$ & 0.8548 & Manipogo & 24.364 & $42.93^{+0.84}_{-0.56}$ \\
\hline
\end{tabular}
\end{table*}

\end{document}